# Is there negative social influence? Disentangling effects of dissimilarity and disliking on opinion shifts

KÁROLY TAKÁCS

*MTA TK "Lendület" Research Center for Educational and Network Studies (RECENS) and Corvinus University of Budapest*

ANDREAS FLACHE

*ICS / University of Groningen*

MICHAEL MÄS

*ETH Zürich*

*Abstract*

*Empirical studies are inconclusive about the underlying mechanisms that shape the interrelated dynamics of opinions and interpersonal attraction. There is strong evidence that others whom are liked have a positive influence on opinions and similarities induce attraction (homophily). We know less about "negative" mechanisms concerning whether disliked others induce shifts away from consensus (negative influence), whether large differences (dissimilarity) generate distancing, and whether dissimilarities induce disliking (heterophobia). This study tests discriminating hypotheses about the presence of positive and negative mechanisms in controlled experiments involving dyadic interactions. Results confirm the presence of homophily, do not support the existence of negative social influence, and show a robust positive linear relationship between opinion distance and opinion shifts. This implies that contact might provide the largest push towards consensus in case of large initial differences.*

# INTRODUCTION

Social influence is a powerful force that is assimilative and fosters opinion convergence in groups (e.g., Festinger, Schachter, and Back 1950; Katz and Lazarsfeld 1955; Fishbein and Ajzen 1975; Nowak, Szamrej, and Latané 1990). People have been shown to assimilate their views to real or perceived opinions of others, even if they privately disagree (e.g., Asch 1956; Cialdini and Petty 1981; Kitts 2003; Willer, Kuwabara, and Macy 2009; Mason, Conrey, and Smith 2007 for a review). Yet, neither small groups nor organizations, neighborhoods, or society at large exhibit perfect consensus, as examples from group discussion experiments as well as studies of political, social and cultural views illustrate (e.g., Huguet, Latané, and Bourgeois 1998; Mark 2003; Glaeser and Ward 2006). Influence dynamics may even result in gradually increasing disagreement. This is suggested by studies of college students (Feldman and Newcomb 1969), international work teams (Early and Mosakowski 2000), and controversial issues in the public debate (Evans 2003; Abramowitz and Saunders 2008; Levendusky 2009).

Persistent diversity is not easy to reconcile with interpersonal influence (cf. e.g., Flache and Macy 2011a). Accordingly, a range of further possible mechanisms and social conditions have been proposed by theories of opinion differentiation. Structural explanations highlight heterogeneity in people's interests or social backgrounds, the impact of opinion leaders and media, or social or geographical boundaries in society that inhibit consensus formation (Hegselmann and Krause 2000; Mark, 2003; Watts and Dodds 2007). In this paper, we examine an additional explanation that origins in different traditions in the classical social psychological literature. Balance theory (Heider 1946), cognitive dissonance theory (Festinger 1957), and social judgment theory (Sherif and Hovland 1961) all justify that positive and negative social influence must be differentiated. They imply that people strive for agreement with a person who is close and liked and for disagreement with persons who are distant and disliked. A strive for cognitive balance is consistent with a motivation to avoid cognitive dissonance: individuals form their opinions in a way to maintain a consistent system of beliefs and attitudes, that is they strive for agreement with a person who is close and liked and for disagreement with persons who are distant and disliked; moreover, they evaluate similar others positively and evaluate dissimilar others negatively.

Recent – mostly formalized - theories on opinion dynamics have accepted the differentiation between positive and negative interpersonal influence (e.g., Macy et al. 2003; Mark 2003; Jager and Amblard 2005; Kitts 2006; Salzarulo 2006; Baldassarri and Bearman 2007; Fent, Groeber, and Schweitzer 2007; Mason, Conrey, and Smith 2007; Flache and Mäs 2008; Flache and Macy 2011a). As a result, their predictions differ from earlier formal theories of social influence that only used positive influence and concluded that consensus in the group is inevitable, unless some subset of group members is entirely cut off from interaction (French 1956; Harary 1959; Abelson and Bernstein 1963; Abelson 1964; DeGroot 1974; Berger 1981; Wagner 1982; Smith and Conrey 2007: 95). Other models considered that individuals may retain some residue of their original positions no matter how large the influence of others (Friedkin and Johnsen 1990; 1999; 2011), but even in these models the resulting diversity is very limited in connected networks (cf. Friedkin 2001). In models that combine positive with negative influence this is different. If at the outset sufficiently many pairs of individuals experience negative social relationships (e.g., mutual disliking), then they distance themselves from each other and shift their opinions at the same time towards positively evaluated group members. The emergent tendency is bi-polarization: divergence towards opposite extreme opinions. Bi-polarization is amplified if – as some models assume – positive and negative influence are combined in a self-reinforcing dynamic with two fundamental micro level mechanisms derived from social-psychological research. These are *homophily* (Lazarsfeld and Merton 1954; Kandel 1978; Platow, Mills, Morrison 2000; McPherson, Smith-Lovin, Cook 2001), modeled as the tendency to like others who are perceived as similar; and *heterophobia*, the assumption that individuals *dislike* dissimilar others (Pilkington and Lydon 1997; Macy et al. 2003; Rydgren 2004; Flache and Mäs 2008; Berger and Dijkstra 2013).

Given that the assumption of negative influence occupies a prominent place in the classical social psychology literature and in contemporary theories of opinion differentiation, it is of great importance to base this assumption on convincing empirical evidence. Most of the empirical research to which theoreticians refer stems from studies of interpersonal power and influence conducted between about 1950 and 1970 (e.g., Mazen and Leventhal 1972), from more recent studies based on social categorization theory (e.g., Hogg, Turner, and Davidson 1990) and on majority and minority influence (e.g., Mucchi-Faina and Cicoletti 2006; Mucchi-Faina and Pagliaro 2008). The evidence they find for negative social influence, however, is inconclusive.

Furthermore, previous empirical studies suffered from the difficulty of differentiating between the underlying mechanisms. Opinion distancing could occur as a result of *dissimilarity* to the source on one hand, and as a result of *disliking* the source on the other. But when dissimilarity and disliking are causally interrelated, as suggested by heterophobia, it is difficult to distinguish in field studies whether changes of opinions are due to one or the other or both. The lack of conclusive evidence and the methodological difficulties in identifying conditions for negative influence empirically have led us to conduct a series of laboratory experiments that test determinants of opinion change when subjects are exposed to interpersonal influence in a highly controlled computerized setting. Our experiments vary dissimilarity between the source and the self systematically, controlling for simultaneous effects of liking and disliking in a longitudinal design. With this, we test predictions derived from theories of negative influence against competing theories that focus exclusively on positive influence and we distinguish between different explanatory mechanisms.

# NEGATIVE SOCIAL INFLUENCE: EVIDENCE AND COMPETING PROPOSITIONS

*Earlier evidence*
Experimental tests have hitherto not provided unequivocal evidence in support of negative influence. In laboratory experiments in the social categorization tradition, researchers typically informed participants about the opinions of members of fictitious in- and out-groups and then measured pre-test – post-test opinion shifts. It was expected that opinions of out-group members should induce negative influence. Yet, many studies did not find increasing differences between in- and out-group opinions at all (Lemaine 1975; Hogg et al. 1990; Krizan and Baron 2007), but only expectations of subjects about the opinion changes of other group members were found to be affected. Some experiments, in which social categorization processes were dominant, found conformity pressures to in-group members only (e.g., Abrams et al. 1990). Furthermore, some experimental designs did not allow disentangling positive influence from the in-group and negative influence from the out-group in the explanation of opinion shifts (e.g., van Knippenberg and Wilke 1988; van Knippenberg et al. 1990; Hogg et al. 1990). In these studies, participants were exposed to in-group members who held opinions relatively similar to their own, but some held more extreme opinions than the subject. Out-group members always had opinions distinct from those of the participants. Researchers found opinion changes away from the out-group opinion (Mackie 1986), but it is not clear whether these were caused by negative influence from the out-group or positive influence from more extreme in-group members.

Moreover, earlier experiments could not differentiate whether opinion distancing occurred as a result of *dissimilarity* or *disliking*. Dissimilarity was often measured as *perceived* dissimilarity between the self and the source of influence along other (e.g., Mackie 1986; Platow, Mills, Morrison 2000), sometimes possibly along irrelevant dimensions. Furthermore, field experiments that extended over a longer time span did not control for general opinion drifts that occurred naturally (e.g., Sampson and Insko 1964; Mazen and Leventhal 1972).

*Positive social influence*
Next, we formulate hypotheses about all mechanisms more precisely and contrast predictions about negative influence to predictions derived from theories of positive social influence. We begin with predictions about positive influence. The "null-model" of interpersonal social influence is that there is always positive influence in interpersonal interaction. More precisely, when an individual is exposed to another opinion, he or she will shift his or her opinion towards the source, regardless of the level of (dis)similarity or (dis)liking. This resonates with experimental evidence about the effect of *mere exposure* to a stimulus, which has been found to enhance an attitude of an individual toward it (Zajonc 1968; Bornstein 1989; Harmon-Jones and Allen 2001).

A *positive opinion-shift* that is *linear* in the original opinion-distance corresponds to the earliest and simplest formal models of social influence (e.g., French 1956; DeGroot 1974; Berger 1981) that assume that individual opinions are updated as a weighted average of the opinions of relevant others. The larger the previous disagreement between the target and the source, the larger is thus the absolute shift of the target's opinion towards the source (*positive influence hypothesis*) according to these theories. In these models, weights are exogenously given for each source-target pair and operationalize the strength of influence in that pair.

Mere exposure and the positive influence hypothesis neglect that opinion distance may also affect liking of the source. Four decades of research in the attraction paradigm showed that the larger the similarity, the larger is also the liking of the source (Byrne 1971; 1997). It has also been argued that similarity increases the perceived relevance of the source (e.g., Stotland, Zander, and Natsoulas 1961; Sherif and Hovland 1961; Burnstein, Stotland, and Zander, 1961; Brock 1965; Hass 1981), which in turn has been linked to attraction to and convergence towards the opinion of the source (e.g., Mazen and Leventhal 1972). Theories of positive influence link differences in perceived similarity to liking, but assume that a lack of similarity and liking does not evoke negative influence. Similarly, formal models explain stable opinion differentiation assuming that interpersonal influence only occurs as long as opinion dissimilarity between the source and the target does not exceed a critical level (Carley 1991; Axelrod 1997; Mark 1998; 2003; Hegselmann and Krause 2002, but see also Mäs, Flache, and Helbing 2010 on the fragility of these explanations).

These models imply that, all other things being equal, the more similar opinions are initially, the more the target should move her opinion towards the source. But the smaller the initial disagreement, the less room there is on the opinion scale for a shift towards the opinion of the source. The interplay of these two opposing effects can be expected to generate an inverted U-shaped effect of the magnitude of initial disagreement on the magnitude of the positive opinion shift towards the source (*similarity hypothesis*). Subjects shift their opinion towards the source, but the magnitude of the shift should increase at a decreasing rate in distance, and even decline beyond some tipping point. The tipping point is defined by the relative steepness of the decline of liking in initial disagreement compared to the increase of the room for a positive shift that comes with more disagreement. Figure 1 illustrates the resulting effect schematically, alongside with competing predictions discussed further below. The horizontal axis of the figure shows the extent of initial opinion-disagreement. The vertical axis charts the predicted direction and magnitude of opinion change of the source. Positive values indicate a shift towards the source's opinion. Further below we introduce hypotheses predicting negative influence, indicated in the figure by negative values that show a shift away from the source. The magnitude indicates how much in terms of the opinion scale the shift reduces or increases the prior disagreement. Figure 1 displays in addition the logical constraints imposed by the boundaries of the opinion scale, indicated by the straight lines above and below the range within which positive or negative opinion shifts can logically fall.

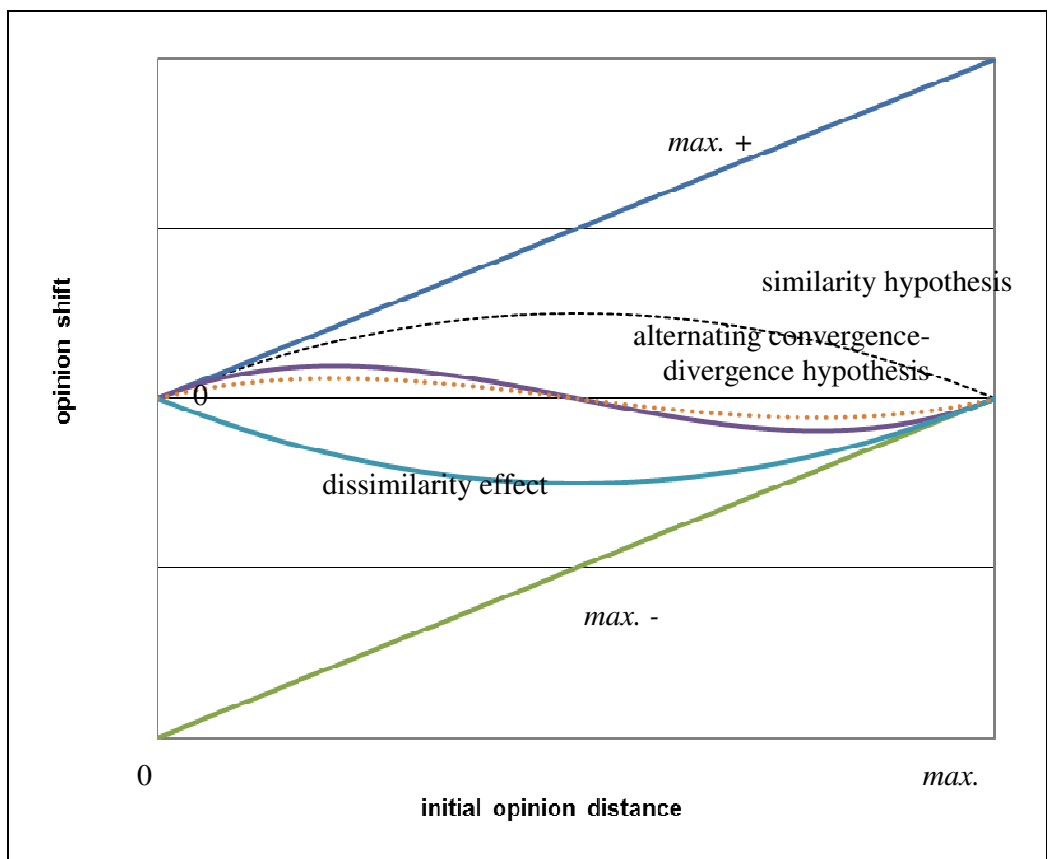

**Figure 1.** Competing hypotheses about the effect of initial opinion distance on opinion shift.
*Notes*: The alternating convergence-divergence hypothesis is depicted with two different parameter specifications. A mathematical justification for the figure can be found in the Supporting Material.

The effect of similarity on opinion shifts can be supplemented or reinforced by the mediating variable of attraction towards the source. Byrne's (1971; 1997) work on the attraction paradigm, has in particular specified the effects of similarity on attraction as a *linear and causal* relationship (Byrne 1971; 1997), hence attraction is updated as a consequence of opinion similarity. In turn, attraction and similarity could both have direct effects on opinion shifts. We test this possibility in experimental treatments where attraction is manipulated independently from opinion-distance.

*Negative influence*

The starting point of this study is that opinions do not necessarily change *towards* the source and evaluations do not necessarily change *in favor* of the source in interpersonal influence. *Negative influence* occurs when individuals change their opinions in order to increase opinion differences. According to the operationalized version of dissonance theory, a subject would reduce dissonance created by the dissimilarity of opinions (or discrepancy) by moving his opinion toward an attractive source (Festinger and Aronson 1960; Aronson, Turner, and Carlsmith 1963). For negatively evaluated sources, however, balance can be restored by increasing dissimilarity with the perceived opinion of the source (Abelson 1964; Mazen and Leventhal 1972; Hogg et al. 1990). Research by Sampson and Insko (1964), for example, suggested that individuals tend to make judgments both in accord with liked others and contrary to disliked others (see also Schwartz and Ames, 1977).

This is mirrored by research on persuasion. Beyond a certain point, additional discrepancy between a position recommended by the source and the initial position of the target may actually decrease persuasion (see Aronson et al. 1963; Abelson 1964; Hass 1981). Other studies have found that a dissimilar communicator may even evoke a "boomerang effect" (Hovland, Harvey, and Sherif 1957) where information from dissimilar others causes inverted attitude change (Sherif and Hovland 1961:Chapter 7; Perloff 1993).

Discrepancy could also have an indirect effect on opinion dynamics through the derogation of the source or rejection. *Heterophobia* is the counterpart of homophily and describes the negative evaluation of others who are dissimilar (Pilkington and Lydon 1997; Flache and Mäs 2008). There is empirical evidence that people might be *socially rejected* (disliked) if they do not conform to the beliefs and opinions of their social group (Festinger and Thibaut 1951; Festinger et al. 1952). In the *attraction paradigm*, Byrne's law specifies a positive linear relationship between similarity and attraction, which can be described by an empirically derived slope and intercept (Byrne 1997). From a positive intercept, it has been concluded that even for a maximum level of dissimilarity, attraction and not repulsion emerges (e.g., Byrne, Clore and Smeaton 1986). This conclusion, however, is wrong, because in these studies, both attraction *and dislike* were measured in the positive domain. It is more appropriate to select the mid-scale of measurement as a borderline in these studies: when the intercept falls below the mid-scale, there is evidence for the dissimilarity-repulsion effect (Rosenbaum 1986; Smeaton, Byrne and Murnen 1989; Chen and Kenrick 2002).

Theories that link dissimilarity to negative opinion change imply that, all other things being equal, beyond a certain critical level of dissimilarity more disagreement induces a larger shift away from the source. However, like with positive influence, the magnitude of a negative opinion shift is logically constrained by the opinion interval. The more disagreement, the less room is left for even more disagreement. In absolute terms this implies a *dissimilarity effect*, which is the mirror image of the non-linear similarity hypothesis for positive influence. When initial opinion-disagreement exceeds a critical level, then further increase of initial disagreement should first induce an increasingly larger negative opinion shift away from the source's opinion. Beyond a tipping point generated by the interplay of increasing need of distancing and decreasing room for disagreement, more initial disagreement should reduce the magnitude of negative opinion shift.

Certain formal theories combine negative influence with positive influence (e.g., Macy et al. 2003; Flache and Macy 2011b; Skvoretz, 2013). More similarity induces a larger relative positive opinion shift when initial disagreement is relatively low. But beyond a critical level of disagreement, less similarity induces a relatively larger negative opinion shift. Due to the constraints of opinion distances, this implies that the non-linear relationships of the similarity hypothesis and of the dissimilarity effect should obtain in separate regions of the scale of initial opinion disagreement. This combination generates a wave-shaped pattern. For low initial opinion-disagreement, more disagreement first increases and then – beyond a tipping point – decreases the magnitude of a positive opinion shift. When initial opinion distance exceeds a critical level, the direction of opinion shift becomes negative. Beyond this point, further increases in initial opinion distance first increase the magnitude of the negative shift and then reduce it. This wave shaped pattern defines the *alternating convergence-divergence hypothesis* that we test in our experiment. We have also elaborated a formal derivation of this wave shaped pattern from the underlying assumptions of the related theories. This can be found in the Supporting Material of this paper.

To give an overview, Figure 1 below summarizes the competing hypotheses for the effects of initial opinion disagreement on the direction and magnitude of the resulting opinion shift of the target of influence.

*Direction and magnitude of opinion changes and attraction: dependent variables*
We devised two computerized experiments to test competing predictions about effects of (dis)similarity and (dis)liking in interpersonal influence. The main dependent variable of our analyses captures a participant's opinion shift after having received information about the opinion of another person (the source). To quantify both direction and magnitude of influence, opinions were always presented and measured on a bounded metric opinion scale ranging from zero to one hundred. The dependent variable opinion shift adopts a positive value if the participant's opinion became more similar to that of the source. Negative values indicate an opinion shift away from the opinion of the source. The magnitude of influence measures in both cases how much the initial opinion disagreement was reduced or increased by the opinion shift. The experiments focused on disentangling effects of two main independent variables: (dis)similarity and (dis)liking on opinion shift. The baseline condition was that subjects' only information was the source's opinion about the issue at stake, such that perceived (dis)similarity could relate only to the extent of prior (dis)agreement.

Our second dependent variable is the attraction rating of the source by the subject. According to the *homophily hypothesis*, a larger extent of similarity implies more liking (higher attraction rates). In contrast, the *heterophobia hypothesis* states that a larger extent of disagreement (dissimilarity) contributes to more disliking. We test both hypotheses in the experimental conditions where we assured that only opinion similarities could contribute to the formation of attraction ratings.

The methodological problem previous research has faced that disagreement has not been considered independently from disliking. We solved this problem by devising a method for introducing a variation in liking independently from the level of opinion disagreement between subjects. Inspired by social balance theory and social categorization theory, we induced negative evaluation of the source of influence by providing subjects with truthful (anonymous) triggers. With this, we could explore the effect of disliking that is not due to dissimilarity on opinion shift. According to the *disliking-divergence hypothesis*, strong (independently manipulated) disliking induces a negative opinion shift of the target of influence away from the source's opinion.

## GENERAL METHOD

Two experiments were conducted. Disliking was not manipulated independently in Experiment 1, but it was in Experiment 2. Both studies were designed with the aim to avoid potential shortcomings of previous research with four core design-features. First, participants were informed about the opinion of only one other participant of the experiment at a time, to assure that opinion shifts were not caused by multiple and potentially conflicting sources of influence. Second, to be able to statistically control for general trends in opinions, we measured participants' opinions before and after being exposed to the opinions of others without any intermediate exposure to other sources of influence. Third, the experiments were designed to avoid that subjects would perceive the situation in in- and out-group terms. Fourth, subjects could not themselves select interaction partners and interacted with each other in a controlled manner. Subjects were aware that their financial compensation was not based on their decisions. In this way, alternative explanations of influence and attraction dynamics that are based on financial motives, on need for consensus, on endogenous interaction dynamics, on argumentation, on persuasive power, on source credibility (Clark, Evans, Wegener 2011), or on perceived threat (cf. Murray and Schaller 2012) could be excluded.

Subjects had to give their opinions on 31 issues (out of which 20 were used in Experiment 2, see Appendix). Issues were selected in multiple pilot studies from an initial list of 83 issues. The selection was based on perceived clarity, sufficiently *high* and sufficiently *similar salience* for participants (see Table A1); and certain desired distribution characteristics in answers (high variance of initial opinions and no concentration on round numbers). All opinions were measured on a 101 point "percentage" scale. Hence, opinion positions were discrete, but small changes could also be detected. We used this scale, because previous opinion dynamic experiments with a much lower number of answer categories found very little change (e.g., Huguet et al. 1998). Issues had furthermore no relation to objective knowledge (cf. Festinger 1950) or to topics for which social desirability effects could be expected to bias responses. Attraction of the subject to the source was measured on a positive scale from 0 to 100, where zero was labeled as "very much disliking" and one-hundred as "very much liking". The question was phrased in Dutch alike the formulation of Byrne (1971:427): "We would like to know your feelings about how much would you probably like this person."

## EXPERIMENT 1

*Procedure*. The experiment took place at the University of Groningen and lasted 45 minutes per session. Participants were randomly seated in cubicles. Subjects responded to a web-based questionnaire. They were first asked about their opinion on and subjectively perceived importance of 31 issues. Then for every subject an issue was selected and "pairs" were formed. Subjects were exposed to the opinion of the source they were paired with on the selected issue (*first stimulus*). This opinion was drawn from a pilot or an earlier session, which allowed using a web-based questionnaire without real online interactions.[1]

After the *first stimulus*, participants rated how much they liked the source (the other person) and gave their own opinion again. This is followed by a *second stimulus* and repeated measurement of liking and opinions.[2]

Each subject participated in 7 to 9 "pairs". Per pair, a new issue was chosen. The order of issues followed a systematic selection procedure that avoided repetition and spread issues evenly across different positions in the sequence in which subjects encountered the issues. Finally, subjects completed a questionnaire asking background data and motivations during the experiment.

---

[1] Subjects were reminded that all participants they are interacting with are real people who *possibly* took part in an earlier session. After the sessions, subjects have received an e-mail about the aim of the experiment and how interaction pairs were formed. This contained the following text: "…Consequently, participants were matched with other subjects, who in fact were from earlier sessions. This was due to technical reasons. (By using different software, we will be able to achieve in the following experiments that every participant is matched with another participant in the same session.)"

[2] Our conclusions were not altered when we analyzed the impact of the second stimulus. The second stimulus had a much lower impact than the first. Details about the subsequent procedure and results are available from the authors upon request.

*Subjects.* Subjects (N=89) were first and second year students of sociology at the University of Groningen who participated as study requirement. They were involved in multiple matches, therefore we obtained 678 observations. N=617 cases were included in the analysis, pilot subjects[3] and incomplete cases were excluded.

RESULTS

*Effects of initial opinion disagreement on opinion changes*
The *positive influence hypothesis,* the *similarity hypothesis* and the *alternating convergence-divergence hypothesis* provide three competing expectations about the effect that the magnitude of the initial opinion-disagreement at the first stimulus has on the magnitude and direction of the subsequent opinion change of the subject. Figure 2 presents a descriptive analysis of the observed association. The horizontal and vertical axes of the figure are defined like in figure 1: the opinion distance between the initially measured opinion of the subject and the opinion of the source to which the subject was exposed at the first stimulus is shown on the horizontal axis and the change of distance after the first stimulus is on the vertical axis. For opinion shifts, positive values indicate by how many scale points the subject shifted towards the opinion of the source and negative values indicate by how many scale points the subject shifted away from the source's opinion.[4]

The curve fitted to the scatterplot in figure 2 shows the average association between the two variables across the 617 observations. It exhibits neither of the non-linear patterns predicted by the *similarity* or the *alternating convergence-divergence hypotheses*. Instead, it seems to be most consistent with the *positive influence hypothesis*. On average, the larger the initial opinion disagreement, the more the subject shifts *towards* the opinion of the source.

In addition, we estimated random-intercept multilevel regression models (Snijders and Bosker, 1999; Raudenbush, Bryk, and Congdon, 2004) of opinion change after the first stimulus, controlling for the nestedness of opinion measurements in subjects. The regression models also included subject-level control variables. Most importantly, linear, quadratic and cubic terms for initial opinion-disagreement tested for the competing patterns predicted by the *positive influence hypothesis,* the *similarity hypothesis* and the *alternating convergence-divergence hypothesis*, respectively.

---

[3] Pilot subjects were excluded because they had artificial partners.
[4] Opinion shifts are calculated as the difference between the original absolute distance to the source $x_1=|o_{i1}-o_{j1}|$ and the new absolute distance to the initial opinion of the source $|o_{i2}-o_{j1}|$ and $|o_{iF}-o_{j1}|$, where $o_{i1}$ is the original opinion of the subject, $o_{j1}$ is the initial opinion of the source, $o_{i2}$ is the opinion of the subject after the first stimulus, and $o_{iF}$ is the final opinion of the subject.

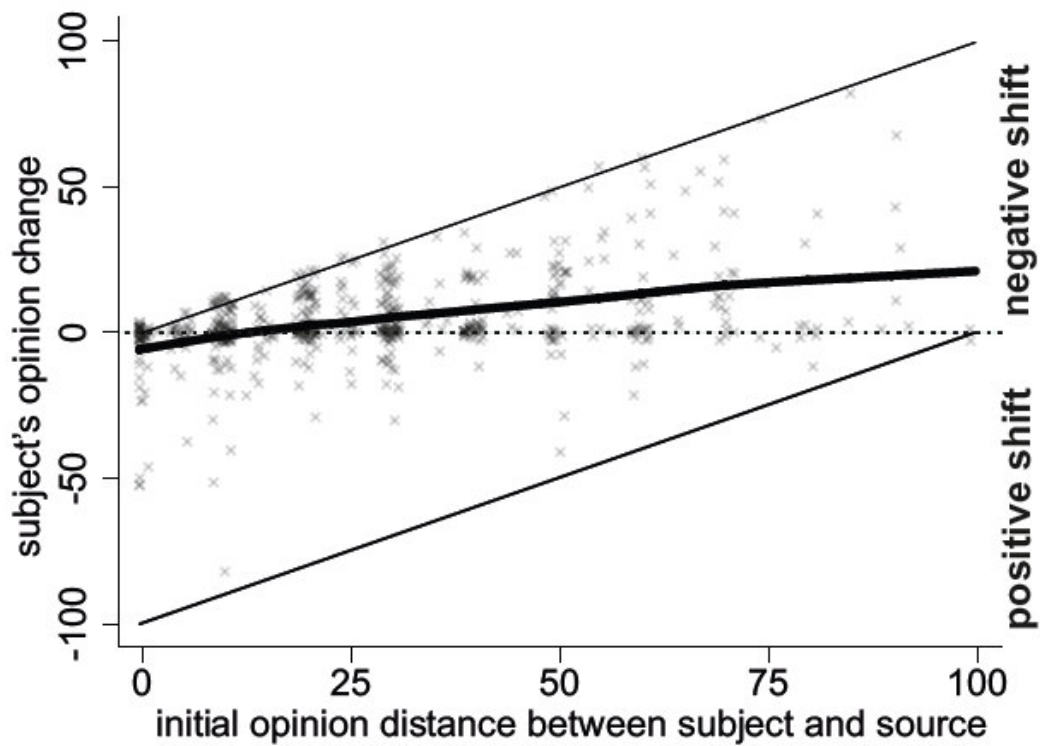

**Figure 2**: Opinion shifts in Experiment 1 by initial opinion distance with plotted lowess smoother: No support for a *dissimilarity effect*

Table 1 shows results for three models that successively add theoretically meaningful control variables. Based on previous work, we controlled whether subjects' resistance to influence were correlated with gender (Guadagno and Cialdini, 2002), study and labor market experience. In addition, model 3 in Table 1 incorporates salience of the selected issue for the subject[5]. The analyses confirm the descriptive results. The relationship between initial opinion-disagreement and opinion change is best approximated by a positively sloped straight line, as predicted by the positive influence hypothesis (Table 1). Non-linear terms for distance are insignificant if included, which refutes both competing hypotheses. These conclusions are not altered when we enter control variables in the analysis. For instance, extremism did not have a significant effect when included in the models as some earlier work would suggest (e.g., Sherif and Hovland 1961; Haslam and Turner 1995).

---

[5] Control variables did not have missing values except the variable "Works" for one subject. This single missing case has been replaced by the mean (0.54).

Interestingly, all intercepts of the models in Table 1 are negative and significant, suggesting that there was a baseline differentiation effect, which is just the opposite of *mere exposure*.

**Table 1:** Results of multilevel regression of opinion change after the first stimulus

| Independent variables | Model 1 | Model 2 | Model 3 |
|---|---|---|---|
| *FIXED EFFECTS* | | | |
| Intercept | -4.35 (1.32)** | -4.40 (1.48)** | -5.71 (2.53)* |
| *Case-level variables* | | | |
| Distance | 0.34 (0.09)*** | 0.35 (0.18)* | 0.34 (0.17)* |
| Distance$^2$/100 | -0.09 (0.14) | -0.12 (0.57) | -0.08 (0.56) |
| Distance$^3$/10000 | | 0.02 (0.47) | -0.01 (0.47) |
| Salience of issue | | | -0.79 (0.78) |
| *Subject-level variables* | | | |
| Gender (female=1) | | | 1.72 (1.47) |
| Year of study | | | 0.75 (0.84) |
| Works (yes=1) | | | 1.48 (1.50) |
| *RANDOM EFFECTS* | | | |
| intercept var. $\mu_0$ | 21.5*** | 21.5*** | 20.9*** |
| level-1 $\sigma^2$ | 159.9 | 160.2 | 160.2 |
| Model deviance | 4943.556251 | 4942.02453 | 4928.76419 |

*Notes*: N=617 cases for 89 subjects. Table shows restricted maximum likelihood HLM2 model estimates obtained in HLM 6. Numbers in parentheses are *robust* standard errors. These are larger than conventional standard errors and have been used as the residual analysis indicated violations of multivariate normality. (Cases with missing values on the main variables have been excluded from the analyses.) All parameter estimates of distance effects are in the meaningful range. For the variance of the random intercept, the *p*-value is obtained from a $\chi^2$-test. The improvement of model deviance between Model 2 and 1 is not significant ($\chi^2(1)=1.53$, $p=0.22$), unlike the improvement between Model 3 and 2 ($\chi^2(4)==13.26$, $p=0.9899$).
\* significant at the 5% level; ** significant at the 1% level; *** significant at the 0.1% level.

*Effects of initial opinion disagreement on liking and disliking*
The average rating of the source by the subject was 61.3 (*SD*=19.1) on an attraction scale that ranged from 0 to 100, suggesting that ratings were rather positive on average. However, 14.3% of the attraction ratings were below the middle of the attraction scale, indicating an unfavorable evaluation. Figure 3 shows a descriptive analysis of the association between initial opinion disagreement and attraction of rating of the source by the subject. The figure depicts a scatterplot of the 617 observations of attraction ratings provided by the 89 participants.

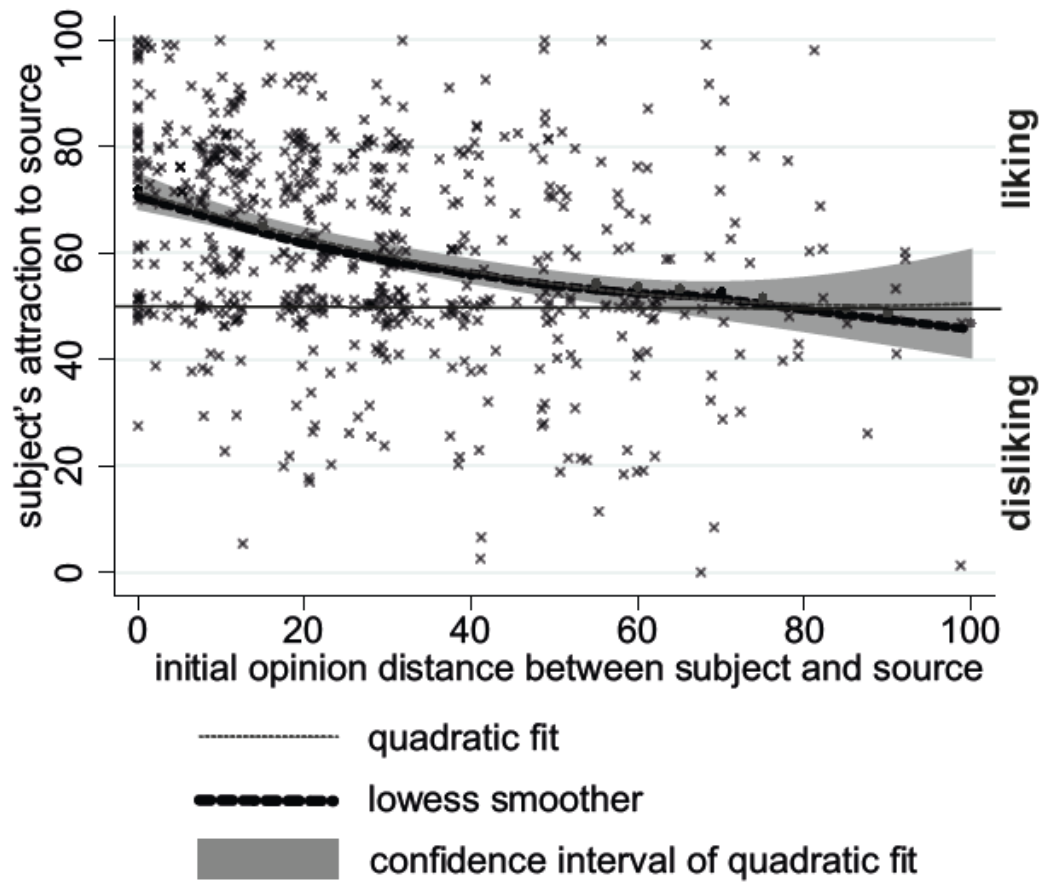

**Figure 3**: Scatter plot of relationship between initial opinion disagreement and attraction measured after the first stimulus in Experiment 1

The average association of the two variables in figure 3 is indicated by the declining quadratic function that was fitted to the data with its corresponding confidence interval (gray area). The direction of the association is consistent with *homophily*. For big initial opinion distances, the fitted line does not significantly fall below the midpoint of the attraction scale, suggesting that even big opinion distances failed to create feelings of disliking (*heterophobia*).

To obtain a conclusive test, we estimated a multi-level regression model that controlled for the nesting of attraction ratings in participants and for individual-level control variables. The results reported in table 2 provide further support for a negative relationship between initial opinion-disagreement and attraction scores. Model 2 supports in addition that the estimated relationship is curvilinear. The combination of the significant negative linear and positive quadratic effect of initial disagreement indicates that attraction rates decrease by opinion distance, but for large distances there is a slight increase. Most importantly, the estimated curve stays above the neutral attraction score of 50. This pattern supports the *homophily hypothesis*, but it is inconsistent with the *heterophobia* hypothesis. Including control variables did not affect this conclusion, but revealed that female subjects gave higher attraction scores on average.

**Table 2:** Results of multilevel regression of attraction ratings after first stimulus

| Independent variables | Model 1 | Model 2 |
| --- | --- | --- |
| *FIXED EFFECTS* | | |
| Intercept | 69.81 (1.67)*** | 75.11 (3.71)*** |
| *case-level variables* | | |
| Distance | -0.29 (0.04)*** | -0.53 (0.11)*** |
| Distance$^2$/100 | | 0.33 (0.12)** |
| Salience | | -1.43 (0.94) |
| *Subject-level variables* | | |
| Gender (female=1) | | 7.30 (2.58)** |
| Works (yes=1) | | -3.72 (2.61) |
| Year of study | | -0.69 (1.19) |
| *RANDOM EFFECTS* | | |
| intercept var. $\mu_0$ | 109.73*** | 103.29*** |
| level-1 $\sigma^2$ | 223.98 | 219.74 |
| Model deviance | 5216.0 | 5189.0 |

*Notes*: N=617 cases for 89 subjects. Table shows restricted maximum likelihood HLM2 model estimates obtained in HLM 6. Numbers in parentheses are *robust* standard errors. The dependent variable "attraction" was measured on a 0…100 scale. For the variance of the random intercept, the *p*-value is obtained form a $\chi^2$-test. The improvement of model deviance between Model 2 and 1 is significant ($\chi^2(5)=27$, *p*=0.9999).
*** significant at the 0.1% level, ** significant at the 1% level, * significant at the 5% level (two-tailed).

BRIEF DISCUSSION

Experiment 1 supports the *positive social influence* and the *homophily hypotheses*. The study did not support the hypotheses we derived from theories of negative influence. We found a small differentiation effect indicating a baseline tendency to differentiate one's opinion from others, but this tendency was combined with a linear and positive effect of opinion-distance on opinion shifts towards the source.

Experiment 1, however, did not allow us to appropriately test those hypotheses that address effects of disliking on opinion shifts (*disliking-divergence hypothesis*). This was because disliking was not manipulated independently from opinion-disagreement so that a possible causal effect of disliking on opinion change could not be assessed.

# EXPERIMENT 2

*Procedure.* The experiment took place in Groningen and lasted one hour per session. For this experiment, new software was developed that allowed real-time computer-mediated communication between the subjects in a pair.[6] The identity of partners in a pair was never revealed. Participants were invited in groups of 10 and were randomly seated in cubicles. First they were asked about their opinion and subjective importance attached to 20 issues (selected from issues in Experiment 1, see Table 1 in Appendix). Subjects were ensured that they interact with partners present in the laboratory, which was indeed the case.

This led us to include two main design changes in the second experiment. First, we decided to manipulate interpersonal attraction, in order to increase the variance of attraction and provide a causal test. Second, subjects now interacted repeatedly with a real other participant so that both changes of opinions and changes of attraction to the other could be observed and tested. This way, we imposed a stronger and longer manipulation in Experiment 2 that triggered potentially a more intense sense of difference between subjects than the manipulation in Experiment 1.

To test for the effect of disliking on opinion shift (*disliking-divergence hypothesis*), in one treatment, an independent attraction manipulation was used, before the first exposure to the opinion of the other participant ("treatment II"). A difficulty we faced in designing the experiment was to induce disliking experimentally in an ethical and sincere way. From earlier experiments we knew that subjects need to perceive the other subject either as someone who is a bad performer on a certain task, behaves in an odd way, has negative or stigmatized characteristics, belongs to a disliked group, or acts offensively towards the subject. In the experiments of Sampson and Insko (1964), dislike was induced by a long procedure involving a failed cooperation task and an insult from the other party. In the experiments of Schwartz and Ames (1977), a negative referent was offending the subject. Keeping this in mind and in order to have sufficient variation in attraction ratings, we used a combination of three methods. We asked participants about their subject of academic study, they had to select an option in a regular Prisoner's Dilemma task, and to choose between sending a stigmatizing or an overwhelmingly positive message to their interaction partner.[7] Subjects learned that their decisions might be (but not necessarily will be) displayed on the screen of their partner. Next, information about the partner's choices was displayed on the screen if the partner had a different subject of study, defected in the Prisoner's Dilemma, or sent a stigmatizing message. If the partner studied the same subject, cooperated, and sent a friendly message, this screen has been left empty. This manipulation was successful in achieving a variation in the initial attraction scores.[8] For treatment II (N=100), a linear regression of initial attraction yielded estimated effects (standard errors are given between brackets) of 68.76 (3.47)*** (intercept) –18.48 (4.39)*** (*Defected*), –11.86 (4.53)** (*Stigmatizing*) –4.10 (4.16) (*SameFaculty*). The model fit was $R^2$=0.29. Hence, only the difference in study direction was not a significant predictor of initial attraction. The other two manipulations had an additive effect and successfully decreased liking of the partner well below the mid-point of the attraction scale (50) that most likely has been understood as the neutral position.

---

[6] The software was developed by Vincent Hindriksen in Delphi.

[7] The two possible messages were formulated as: 1. "I am a very nice person. I will do all my best to help you and nobody else in this experiment."; 2. "You have to know that I want to do my best in this experiment and I do not care about what you are going to receive."

After the initial measurement of attraction, the opinion of the partner was presented and attraction and opinion were measured again. In addition, subjects could select one from the same list of persuasive messages to be sent to the other person. In the following screen, along with the opinion, the persuasive message that was sent by the partner was displayed, and attraction and opinion were recorded again. In case attraction was not manipulated, the procedure until this point was the same as in Experiment 1 except the attraction measurement at the start. In Experiment 2, however, this last step was repeated once more in both treatments: subjects could select a persuasive message again, they received this message from the partner along with his or her opinion, and they had to rate the partner and give their opinions again.[9]

Subjects were matched with each other based on a complex algorithm developed and programmed by the first author. The algorithm excluded issues with insufficient variation in initial opinions, and was designed to simultaneously maximize the variance of initial opinion distances across pairs, the variance of distances to the opinion of the partner within the individual across matches, and to minimize the inequality of salience within pairs. The iterated algorithm selected 9 issues that provided the best solutions for these criteria and determined a random sequence among these issues. As a consequence, subjects participated in 9 dyadic interactions, each with a different issue and an unknown partner. The experiment was followed by a questionnaire asking background data and motivations during the experiment.

*Subjects.* Subjects (N=110) were students from all faculties of the University of Groningen gathered via board advertisements, lecture announcements, and advertisements in the university newspaper. Participants were paid 8 euro for their participation independently from their choices in the experiment. All subjects had in addition an equal chance to win 200 euro in a lottery. Subjects played 92 issue rounds in 11 sessions (920 valid cases). Data from one session (N=10) has been excluded from the analysis due to missing values of the dependent variable.

---

[8] Attraction was measured the same way as in Experiment 1, on a scale of 0…100.
[9] The further round of persuasion did not alter our conclusions about the main effects.

RESULTS

*Effects on opinion changes*

In treatment II of Experiment 2, attraction was independently manipulated with the aim to induce disliking. As we did not deceive subjects, there was a variation about the information subjects received during the manipulation and consequently also in the initial (dis)liking score. The mean attraction rate in treatment II was 56.67 (*SD*=21.85), which is still above the midscale value, but significantly lower than at the first measurement in treatment I (60.22; *t=2.47*). With more disliking, we expected that negative opinion changes (away from the source) should occur more frequently in treatment II. Table 3 gives a description of the distribution of directions of opinion change across the treatments of the study, broken down into the categories of positive, negative and neutral (no) change. The table shows in addition how the direction of opinion change was associated with the mean initial opinion-disagreement for the cases in the given category. For comparison, distributions are shown both for opinion change after the first stimulus and the aggregated opinion change that had resulted after all stimuli.

**Table 3**: Number of cases for positive, negative, and no opinion shifts in Experiment 2 after the first stimulus and after all stimuli.

| *Treatment I* | after 1st stimulus | Mean initial distance (*SD*) | after all stimuli | Mean initial distance (*SD*) |
|---|---|---|---|---|
| Positive shifts | 132 (34.7%) | 36.8 (21.8) | 206 (54.2%) | 35.1 (21.8) |
| No change | 220 (57.9%) | 23.4 (22.6) | 145 (38.2%) | 19.9 (22.0) |
| Negative shifts | 28 (7.4%) | 22.8 (20.4) | 29 (7.6%) | 18.2 (19.5) |
| Total | 380 | 28.0 (23.0) | 380 | 28.0 (23.0) |
| *Treatment II* | | | | |
| Positive shifts | 155 (33.0%) | 37.7 (22.5) | 253 (53.8%) | 36.1 (22.1) |
| No change | 274 (58.3%) | 26.5 (23.5) | 173 (36.8%) | 22.6 (23.8) |
| Negative shifts | 41 (8.7%) | 17.4 (18.2) | 44 (9.4%) | 17.6 (18.8) |
| Total | 470 | 29.4 (23.6) | 470 | 29.4 (23.6) |

*Notes*: Mean initial opinion distances for the corresponding cases are indicated.

Table 3 shows that there is no discernible difference in the occurrence of negative opinion shifts between treatments I and II. Of all observed opinion changes after the first stimulus 7.4% and 8.7% were negative in treatment I and II, respectively. For the overall change after all stimuli this was 7.6% vs. 9.4%. While these figures indicate slightly more negative influence in treatment II, the average extent of shifts was not statistically different (*d=-0.56*, *t=-0.67* for the first shift, *d=-0.006*, *t=-0.007* after all stimuli). Moreover, comparison of the mean initial opinion distances associated with the different categories shows that negative opinion shifts were associated with smaller initial distances rather than larger ones, contrary to what theories of negative influence suggest. This pattern was further confirmed when we conducted for Experiment 2 a descriptive analysis like the one presented in figure 2 for Experiment 1. When we fitted a curve to the scatterplot of all observed pairs of initial opinion-disagreement and opinion change after the first stimulus, the prevalent pattern was very similar to what we found in Experiment 1 (additional results are available upon request from the authors).

To obtain a controlled test, we estimated random-intercept multilevel regression models of opinion change after the first stimulus separately for treatments. The only difference to the models estimated for Experiment 1 was that we added for treatment II terms that allowed assessing the unique contribution of the independently manipulated initial attraction, as well as its potential interaction with initial disagreement. Table 4 displays results.

Results for treatment I show a somewhat different pattern than we obtained for Experiment 1. In its basic setup, Treatment I did not differ much from the design of Experiment 1 as the opinion of the source was the only stimulus for shifting an opinion. Results are similar in some aspects, but while in Experiment 1 we found a strong positive linear effect of initial opinion disagreement on the magnitude of positive opinion shift, results of treatment I in Experiment 2 lend some support for non-linear effects of initial disagreement. For ease of interpretation, figure 4 below shows the predicted association between initial opinion-disagreement and the resulting opinion shift for different models that we estimated. Both models 1 and 2 in table 4 comprise significant effects of initial disagreement (positive) as well as initial disagreement squared (negative). Moreover, the cubic term of initial disagreement in model 2 has a positive effect that is close to being statistically significant at the $p=0.05$ level. The graphical analysis in figure 4 shows that the estimated curve remotely resembles the wave-shape predicted by the *alternating convergence-divergence hypothesis* (also shown in the figure). Unfortunately, this cannot be interpreted as support of the hypothesis. On closer inspection, we find that the estimated curve falls predominantly into the region of positive opinion shifts. This holds in particular for large initial opinion disagreement, in contradiction with the hypothesis. Moreover, at all levels of initial disagreement the estimated curve does not exhibit a negative slope, thus also contradicting the pattern predicted by the *similarity hypothesis*. Despite the differences with Experiment 1, we conclude that also the results for treatment I in Experiment 2 are best approximated by the positive effect predicted by the *positive influence hypothesis*. This also holds for the results obtained for treatment II, as models 3 and 4 exemplify. While initial opinion disagreement had a significant and positive effect on opinion change, the non-linear effects of initial disagreement were not significant. Moreover, also here the estimated curves fall predominantly into the region of positive opinion shifts (see figure 4). As a further test, we estimated similar models also for the second opinion shift in treatment II, with the same overall result (available upon request).[10]

**Table 4:** Results of multilevel regression of first opinion change in Experiment 2 (change in absolute distance to the initial opinion of the source)

| Independent variables | Treatment I | | Treatment II | |
|---|---|---|---|---|
| *FIXED EFFECTS* | Model 1 | Model 2 | Model 3 | Model 4 |
| Intercept | -3.14 (0.88)*** | -1.19 (1.77) | 1.34 (1.70) | -0.66 (2.16) |
| *case-level variables* | | | | |
| Distance | 0.64 (0.19)*** | 0.63 (0.19)*** | 0.28 (0.13)* | 0.29 (0.13)* |
| Distance$^2$/100 | -1.53 (0.75)* | -1.51 (0.74)* | -0.67 (0.46) | -0.70 (0.46) |
| Distance$^3$/10000 | 1.28 (0.71)[11] | 1.26 (0.70)[12] | 0.49 (0.38) | 0.51 (0.38) |
| Initial attraction | | | -0.07 (0.03)* | -0.07 (0.03)* |
| Attraction * distance | | | 0.0027 (0.0013)* | 0.0027 (0.0013)* |
| Salience of issue | | -0.20 (0.59) | | 0.77 (0.62) |
| *Subject-level variables* | | | | |
| Gender (female=1) | | -0.82 (1.41) | | 1.16 (1.37) |
| Works (yes=1) | | -0.86 (1.36) | | 0.52 (1.10) |
| Master student (yes=1) | | -2.60 (1.28)* | | -0.31 (1.07) |
| *RANDOM EFFECTS* | | | | |
| intercept var. $\mu_0$ | 15.65*** | 15.31*** | 0.85 | 0.46 |
| level-1 $\sigma^2$ | 100.9 | 101.2 | 127.6 | 128.4 |
| Model deviance | 2882.3 | 2869.3 | 3636.6 | 3626.7 |

*Notes*: N=380 cases in treatment I; N=470 cases in treatment II for 100 subjects in each treatment. Missing values (N=9) for the Works variable have been imputed by the overall mean (0.411). The interaction variable Attraction*distance was simply created as multiplying the Attraction (scale: 0…100) and the Distance (scale: 0…+100) variables. Table shows restricted maximum likelihood HLM2 model estimates obtained in HLM 6. Numbers in parentheses are *robust* standard errors. All parameter estimates of distance effects are in the meaningful range.

\* significant at the 5% level; \*\*\* significant at the 0.1% level. For testing random effects $\chi^2$ tests are used.

---

[10] Opinion shifts after the second stimulus were far smaller than shifts after the first stimulus. This is in line with other experimental findings on social influence. Moreover, the smaller shifts after the second stimulus were in the same direction as first shifts and could even be described with the same underlying mechanisms except they were smaller in size. These results clearly demonstrate that a longer interaction is not different qualitatively from a short influence process and it is the first impression that matters the most.
[11] $p$=0.071 for robust standard errors, $p$=0.000 for regular standard errors.
[12] $p$=0.072 for robust standard errors, $p$=0.000 for regular standard errors.

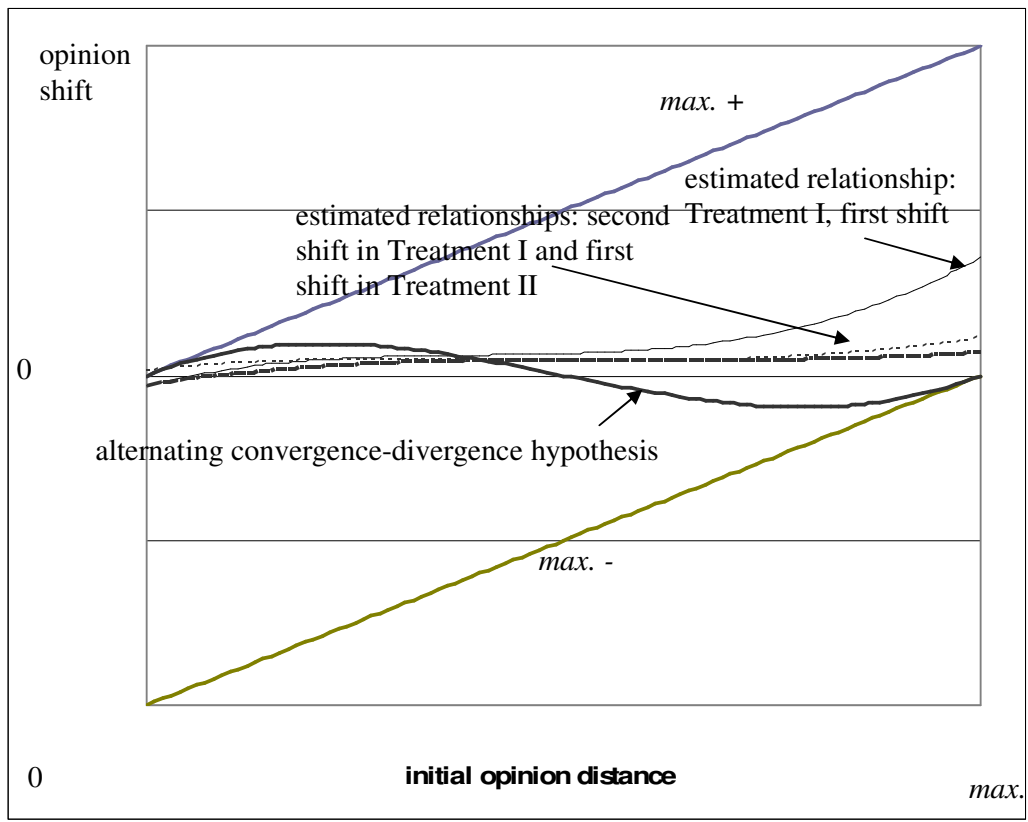

**Figure 4**. The curvilinear relationship between opinion distance and opinion shifts in Experiment 2, based on the parameter estimates of Models 1 and 3 in Table 4 (first shift) and Model 1 for the second shift (available upon request)

The effects of independently manipulated attraction on opinion change after the first stimulus allow testing the *disliking-divergence hypothesis*. Remarkably, we find a significant *negative effect of attraction* on opinion change (models 3 and 4, table 4). This contradicts the hypothesis that postulates that disliking rather than liking reduces and even inverts positive opinion shifts. The main effect of attraction, however, is to be interpreted in combination with the positive interaction effect of initial disagreement and that of initial attraction. For interpretation of the overall effect of disliking on opinion change in treatment II, we compared the average opinion change between observations, for which the subject indicated disliking of the source, to those where liking was indicated, broken down by small vs. large initial opinion disagreement (Figure 5). Results do not support the *disliking-divergence hypothesis*. Overall, average opinion changes were positive in all categories. This pattern of results shows no indication of negative opinion shifts induced by disliking.

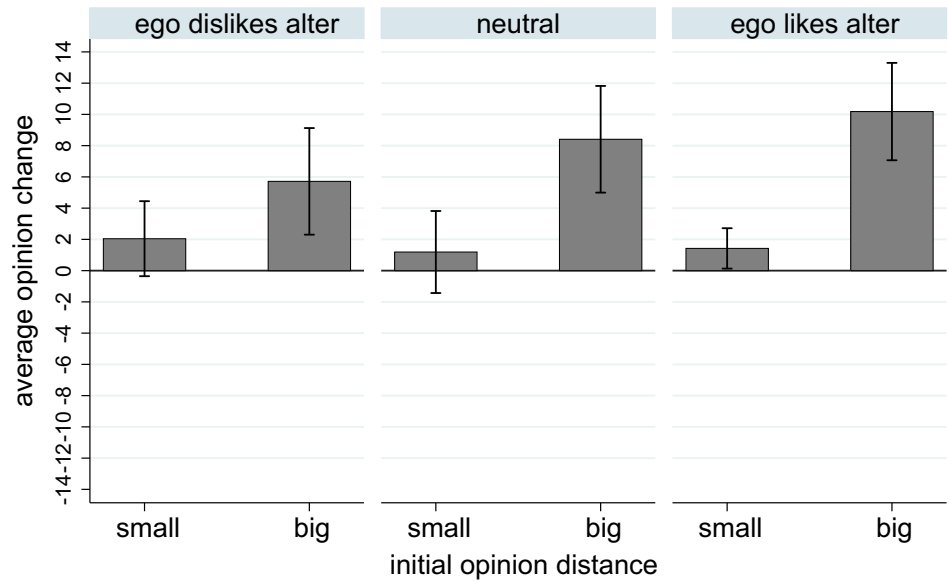

**Figure 5.** Direction and average magnitude of opinion change after first stimulus in treatment II (with attraction manipulation) of Experiment 2, broken down by liking vs. disliking and small vs. large initial distances.
*Note*: small: distance below average, big: distance above average

*Effects of initial opinion disagreement on liking and disliking*
We tested the *homophily hypothesis* and the heterophobia hypothesis with multilevel regression models of subject's attraction towards the partner, which was measured after the first stimulus. Again, the main difference compared to Experiment 1 was that we estimated models separately for treatments I and II, and controlled in treatment II for effects of independently manipulated attraction ratings. Table 5 displays the results.

All models in table 5 show that attraction ratings declined in initial opinion-disagreement. The large intercept terms show that on average, subjects like their partner when initial disagreement is relatively small. For treatment I, we observe that unlike in Experiment 1, estimated liking turns into disliking when initial disagreement exceeds a critical level. For maximum initial differences, predicted ratings drop well below the mid-point of the scale of attraction (50), especially after controlling for background variables in model 2. These results are consistent with the *homophily hypothesis* and lend some support to the *heterophobia hypothesis*. In treatment II, attraction (disliking) was manipulated independently first. Initial attraction ratings have therefore been included in the analysis. It turned out that attraction ratings determined the next measurement to a large extent (see the parameter for initial attraction in models 3 and 4 in table 5). The significant positive intercept in models 3 and 4 indicates a general tendency to evaluate the partner more positively after the first stimulus than before. Still, for large initial distances, estimated attraction ratings are below the midscale value, supporting the *heterophobia hypothesis*.

**Table 5:** Results of multilevel regression of attraction ratings after the first stimulus in Experiment 2.

| Independent variables | Treatment I | | Treatment II | |
|---|---|---|---|---|
| | Model 1 | Model 2 | Model 3 | Model 4 |
| *FIXED EFFECTS* | | | | |
| Intercept | 67.91 | 70.25 | 30.52 | 35.64 |
| | (1.94)*** | (4.82)*** | (3.27)*** | (5.56)*** |
| *case-level variables* | | | | |
| Distance | -0.27 | -0.38 | -0.21 | -0.40 (0.13)** |
| | (0.04)*** | (0.11)*** | (0.04)*** | |
| Distance$^2$/100 | | 0.0015 (0.001) | | 0.0011 (0.001) |
| Initial attraction | | | 0.59 (0.04)*** | 0.53 (0.08)*** |
| Distance * attraction | | | | 0.002 (0.0016) |
| Salience | | -0.67 (1.25) | | 0.76 (1.00) |
| *subject-level variables* | | | | |
| Gender (female=1) | | 1.92 (3.76) | | 1.06 (2.17) |
| Works (yes=1) | | -3.45 (3.34) | | -2.68 (1.68) |
| MA student (yes=1) | | -1.57 (3.17) | | -4.76 (1.57)** |
| *RANDOM EFFECTS* | | | | |
| intercept var. $\mu_0$ | 159.02*** | 163.13*** | 10.80 | 4.20 |
| level-1 $\sigma^2$ | 239.69 | 239.58 | 279.93 | 279.87 |
| Model deviance | 3110.7 | 3101.6 | 3831.7 | 3828.2 |

*Notes*: N=360 cases in treatment I and N=450 cases in treatment II for 90 subjects in each treatment. The dependent variable is the attraction score measured after the first stimulus (first display of the opinion of the pair). Note that this is the first attraction measurement in treatment I and the second measurement in treatment II. Table shows restricted maximum likelihood HLM2 model estimates obtained in HLM 6. Numbers in parentheses are *robust* standard errors. (Cases with missing values have been excluded from the analyses.)

\*\*\* significant at the 0.1% level, \*\* significant at the 1% level, \* significant at the 5% level (two-tailed).

BRIEF DISCUSSION

Experiment 2 was designed to induce disliking with an independent attraction manipulation (treatment II) and to allow assessing causal relations between attraction and opinion shift in this way. Contrary to what theories of negative influence suggest, the results do not support that disliking induces negative influence. We find some evidence that large initial disagreement triggers disliking. Yet, overall, average opinion changes were positive and became more so to the extent that initial disagreement was larger. Again, this result lends most support to the *positive* *influence hypothesis* that larger opinion differences induce a larger positive shift towards the source.

## OVERALL SUMMARY AND CONCLUSION

Negative influence occurs when individuals change their attitudes in order to increase opinion differences to negatively evaluated others. Negative influence has been considered in different classical social psychological theories and has been re-invented recently in models to explain opinion differentiation and bi-polarization; outcomes that are difficult to reconcile with earlier models of opinion formation. Despite this prominent place of negative influence in the classical social psychological literature and in recent theories of opinion dynamics, there is a dearth of reliable empirical evidence for this mechanism at the micro-level of interpersonal influence. To address this, we conducted two experimental studies in a highly controlled computer-mediated setting, in which we systematically varied the initial opinion-disagreement between a subject and a source of opinion, and the liking or disliking of the source independently.

We articulated a number of hypotheses from alternative sets of assumptions on social influence and tested these with the data obtained from the two experimental studies. Overall, results are more supportive for theories of *positive influence* than they are for theories of negative influence. Most importantly, even when attraction was manipulated (disliking in Experiment 2), it did not trigger opinion shift *away* from the opinion of the source, which was predicted by theories of negative influence. The strongest and most general effect we found is a positive linear effect of opinion distance on opinion shifts[13]. A linear effect of opinion distance is in line with classical models of social influence (e.g., French, 1956). This finding implies for models of opinion dynamics that a complex non-linear social influence function might be unnecessary to characterize the relationship between similarity and opinion change. Our results suggest that not only for the sake of simplicity, but also for the sake of realism, model builders should be cautioned against resorting too readily to a more complex assumption than a simple linear influence function.

We found clear support for *homophily* in our experiments, but only mixed evidence for *heterophobia*, which could be evoked by large initial disagreement between a subject and a source of influence. In Experiment 1 we did not find evidence to support the heterophobia hypothesis, but in Experiment 2 we did. Large initial disagreement induced disliking in this experiment, but contrary to theories of negative influence, this was not associated with negative opinion shifts. In short, our findings are only partly consistent with a motivation to reduce cognitive dissonance (Festinger 1957; Festinger and Aronson 1960; Aronson et al. 1963) and with predictions of social judgment theory (Sherif and Hovland, 1961). Individuals might have formed their opinions in a way to build and maintain a consistent system of beliefs and opinions, but this was not strictly differentiated based on liking of the source. Subjects in our experiments tended to shift their opinions towards those of the source, irrespective of their evaluation. This might be because people in general do not easily distance away their opinion from others, even if the other one has an extreme position or is not very much liked. Yet a possible interpretation is that the most distant opinions were the most disturbing for the subjects, but they reduced dissonance only by shifting opinions *towards* the source, and never by *moving away* from or by derogating the source. This might mean that despite the successful manipulations, "liking" and "disliking" in our experiments were not of crucial importance for opinion formation and cognitive balance.

---
[13] Mavrodiev, Tessone, and Schweitzer (2013) recently reported a similar finding for a social influence experiment in which subjects had to guess the right answer to a factual question.

There are several caveats and limitations concerning our studies and the generalizability of our results. Arguably, our artificial laboratory setting with computer-mediated anonymous interaction may have suppressed the emotional processes that in field settings induce disliking and rejection of others' opinions. In case of face-to-face encounters, subjects' visible characteristics, sexual attraction, and facial expressions would be important variables that are difficult to measure and to control for. We could exclude such factors, but the cost may have been that we could not observe some of the negative influence dynamics that might occur in field settings. We cannot claim ex post, however, that our manipulations were insufficiently "weak". These manipulations were not weak for positive social influence as half of the subjects shifted their opinions as a result of receiving information about the opinion of their partner towards the source. Ex ante, we did not want to create an experimental setting that is able to prove the existence of negative influence. We intended to test hypotheses about the existence of the hypothesized effects in simple and standard experimental situations in line with the theoretical assumptions made in important theories.

We have reduced the social influence setting to interaction in dyads, without any concept of groups with which subjects might identify or they might reject as negatively evaluated out-group. This was done on purpose, and could be relaxed in future research.

Another important note is that in roughly half of the cases, subjects did not change their opinions at all. This was partially because subjects were biased towards round numbers on the opinion scale most of the time; more so for attraction scores than for opinions. Furthermore, our analyses indicate that much of the variation in subjects' attraction ratings and opinion changes is not explained by the initial disagreement or initial attraction. Individuals seem to vary on how open they are to positive and negative influence, how persuasive they are and how intolerant they are for large inconsistencies of opinions (e.g., Janis et al. 1959; Hovland and Rosenberg 1960; Rosenberg 1960; Eagley 1981; Clark, Evans, Wegener 2011). While we tried to capture some of this variation with individual level controls, future research could improve upon this and include measures of corresponding personality variables.

Overall, our research provides a highly controlled test of some of the assumptions on negative social influence that have a prominent place in the classical social psychological literature and in recent models of opinion formation. We find only very little evidence that these assumptions serve well to describe the behavior of subjects in the computer-mediated social influence settings of our experiment. We cannot exclude that negative influence may occur under conditions in the laboratory or field that we have not captured in our experiments. We think, however, that our findings point to the need to inspect more carefully by which mechanisms and under what conditions negative influence is a plausible assumption in theories of opinion dynamics.

# REFERENCES

Abelson, R. P. (1964). Mathematical models of the distribution of attitudes under controversy. In N. Frederiksen & H. Gulliksen (Eds.), *Contributions to mathematical psychology* (pp. 142-160). New York: Holt, Rinehart, and Winston.

Abramowitz, A. I., & Saunders, K. L. (2008). Is polarization a myth? *Journal of Politics*, 70, 542-555. DOI: 10.1017/S0022381608080493

Abrams, D., Wetherell, M., Cochrane, S., Hogg, M. A., & Turner, J. C. (1990). Knowing what to think by knowing who you are: Self-categorisation and the nature of norm formation, conformity and group polarization. *British Journal of Social Psychology*, 29, 97-119. DOI: 10.1111/j.2044-8309.1990.tb00892.x

Aronson, E., Turner J. A., & Carlsmith, J. M. (1963). Communicator credibility and communication discrepancy as determinants of opinion change. *Journal of Abnormal and Social Psychology*, 67, 31-36. DOI: 10.1037/h0045513

Asch, S. E. (1956). *Studies of Independence and Conformity*. Washington D.C.: American Psychological Association.

Axelrod, R. (1997). The dissemination of culture. A model with local convergence and global polarization. *Journal of Conflict Resolution*, 41, 203-226. DOI: 10.2307/174371

Baldassarri, D., & Bearman, P. (2007). Dynamics of political polarization. *American Sociological Review*, 72, 784-811. DOI: 10.1177/000312240707200507

Berger, C., & Dijkstra, J. K. (2013). Competition, envy, or snobbism? How popularity and friendships shape antipathy networks of adolescents. *Journal of Research on Adolescence*, 23: 586–595. doi: 10.1111/jora.12048

Berger, R. L. (1981). A necessary and sufficient condition for reaching a consensus using degroot's method. *Journal of the American Statistical Association*, 76, 415–419. DOI: 10.1080/01621459.1981.10477662

Brauer, M., & Judd, C. M. (1996). Group polarization and repeated attitude expressions: A new take on an old topic. *European Review of Social Psychology*, 7, 173-207. DOI: 10.1080/14792779643000010

Brewer, M. B. (1999). The psychology of prejudice: In-group love or out-group hate? *Journal of Social Issues*, 55, 429-444. DOI: 10.1111/0022-4537.00126

Brock, T. C. (1965). Communicator-recipient similarity and decision change. *Journal of Personality and Social Psychology*, 1, 650-654. DOI: 10.1037/h0022081

Burnstein, E., Stotland, E., & Zander, A. (1961). Similarity to a model and self-evaluation. *Journal of Abnormal and Social Psychology,* 62, 257-264. DOI: 10.1037/h0043981

Byrne, D. (1971). *The attraction paradigm*. New York: Academic Press.

Byrne, D. (1997). An overview (and underview) of research and theory within the attraction paradigm. *Journal of Social and Personal Relationships*, 14, 417-431. DOI: 10.1177/0265407597143008

Byrne, D., Clore, G. L., & Smeaton, G. (1986). The attraction hypothesis: Do similar attitudes affect anything? *Journal of Personality and Social Psychology*, 51, 1167-1170. DOI: 10.1037/0022-3514.51.6.1167

Carley, K. (1991). A theory of group stability. *American Sociological Review*, 56, 331-354. DOI:10.2307/2096108

Chen, F. F., & Kenrick, D. T. (2002). Repulsion or attraction? Group membership and assumed attitude similarity. *Journal of Personality and Social Psychology*, 83, 111-125. DOI: 10.1037//0022-3514.83.1.111

Cialdini, R. B., & Petty, R. E. (1981): Anticipatory Opinion Effects. In: Petty, R. E.; Ostrom, T. M., & Brock (eds.): *Cognitive Responses in Persuasion*. Hillsdale, NJ, Lawrence Erlbaum Associates.

Clark, J. K., Evans, A. T., & Wegener, D. T. (2011). Perceptions of source efficacy and persuasion: Multiple mechanisms for source effects on attitudes. *European Journal of Social Psychology*, 41: 596–607. DOI: 10.1002/ejsp.787

DeGroot, M. H. (1974). Reaching a consensus. *Journal of the American Statistical Association*, 69, 118–121. DOI: 10.1080/01621459.1974.10480137

Eagley, A. H. (1981). Recipient characteristics as determinants of responses to persuasion. In R. E. Petty, T. M. Ostrom, & T. C. Brock (Eds.), *Cognitive Responses in Persuasion* (pp. 173-195). Hillsdale: Lawrence Erlbaum Associates.

Evans, J. H. (2003). Have americans' attitudes become more polarized? - An update. *Social Science Quarterly*, 84, 71-90. DOI: 10.1111/1540-6237.8401005

Feldman, K. A., & Newcomb, T. M. (1969). *The impact of college on students*. San Francisco: Jossey-Bass.

Fent, T., Groeber P., & Schweitzer, F. (2007). Coexistence of social norms based on in- and out-group interactions. *Advances in Complex Systems*, 10, 271-286. DOI: 10.1142/S0219525907000970

Festinger, L. (1950). Informal social communication. *Psychological Review*, 57, 271-282. DOI: 10.1037/h0056932

Festinger, L. (1957). *A theory of cognitive dissonance*. Stanford: Stanford University Press.

Festinger, L., & Aronson, E. (1960). The arousal and reduction of dissonance in social contexts. In D. Cartwright & A. Zander (Eds.), *Group dynamics: Research and theory* (pp. 125-136). Evanston: Row and Peterson.

Festinger, L., & Thibaut, J. (1951). Interpersonal communication in small groups. *Journal of Abnormal and Social Psychology*, 46, 92–99. DOI: 10.1037/h0054899

Festinger, L., Gerard, H. B., Hymovitch, B., Kelley, H. H., & Raven, B. (1952). The influence process in the presence of extreme deviates. *Human Relations*, 5, 327–346. DOI: 10.1177/001872675200500402

Festinger, L., Schachter S., & Back, K. (1950). *Social pressures in informal groups.* New York: Harper.

Fishbein, M., & Ajzen, I. (1975). *Belief, Attitude, Intention and Behavior. An Introduction to Theory and Research.* Reading, MA: Addison-Wesley.

Flache, A., & Macy, M. W. (2011a). Local convergence and global diversity: From interpersonal to social influence. *Journal of Conflict Resolution,* 55, 970-995. DOI: 10.1177/0022002711414371

Flache, A., & Macy, M. W. (2011b). Small worlds and cultural polarization. *Journal of Mathematical Sociology*, 35: 146-176. **DOI:**10.1080/0022250X.2010.532261

Flache, A., & Mäs, M. (2008). How to get the timing right: A computation model of the effects of the timing of contacts on team cohesion in demographically diverse teams. *Computational and Mathematical Organization Theory*, 14, 23-51. DOI: 10.1007/s10588-008-9019-1

French, J. R. P. (1956). A formal theory of social power. *Psychological Review*, 63, 181–194. DOI: 10.1037/h0046123

Friedkin, N. E. (2001). Norm formation in social influence networks. *Social Networks*, 23, 167-189. DOI: 10.1016/s0378-8733(01)00036-3

Friedkin, N. E., & Johnsen, E. C. (1990). Social influence and opinions. *Journal of Mathematical Sociology,* 15, 193–205. **DOI:** 10.1080/0022250X.1990.9990069

Friedkin, N. E., & Johnsen, E. C. (1999). Social influence networks and opinion change. *Advances in Group Processes,* 16, 1-29.

Friedkin, N. E., & Johnsen, E. C. (2011). *Social influence network theory*. New York, NY: Cambridge University Press.

Guadagno, R. E., & Cialdini, R. B. (2002). Online persuasion: An examination of gender differences in computer-mediated interpersonal influence. *Group Dynamics: Theory, Research and Practice*, 6, 38-51. DOI: 10.1037/1089-2699.6.1.38

Harary, F. (1959). A criterion for unanimity in French's theory of social power. In D. Cartwright (Ed.), *Studies in social power* (pp. 168-182). Ann Arbor: Institute for Social Research.

Haslam, S. A., & Turner, J. C. (1995). Context-dependent variation in social stereotyping 3: Extremism as a self-categorical basis for polarized judgement. *European Journal of Social Psychology*, 25, 341-371. DOI: 10.1002/ejsp.2420250307

Hass, R. Glen (1981). Effects of source characteristics on cognitive responses and persuasion. In R. E. Petty, T. M. Ostrom, & T. C. Brock (Eds.), *Cognitive responses in persuasion* (pp. 141-172). Hillsdale: Lawrence Erlbaum Associates.

Hegselmann, R., & Krause, U. (2002). Opinion dynamics and bounded confidence: Models, analysis and simulation. *Journal of Artificial Societies and Social Simulation*, 5, Retrieved June 30, 2002, from http://jasss.soc.surrey.ac.uk/5/3/2.html

Heider, F. (1946). Attitudes and cognitive organization. *Journal of Psychology*, 21, 107-112. DOI: 10.1080/00223980.1946.9917275

Hogg, M. A., Turner, J. C., & Davidson, B. (1990). Polarized norms and social frames of reference: A test of the self-categorization theory of group polarization. *Basic and Applied Social Psychology,* 11, 77-100. **DOI:** 10.1207/s15324834basp1101_6

Hovland, C. I., & Rosenberg, M. J. (1960). Summary and further theoretical issues. In: M. J. Rosenberg, I. H. Carl, W. J. McGuire, R. P. Abelson, & J. W. Brehm (Eds.), *Attitude organization and change.* New Haven: Yale University Press.

Hovland, C. I., Harvey, O. J., & Sherif, M. (1957). Assimilation and contrast effects in reactions to communication and attitude-change. *Journal of Abnormal and Social Psychology*, 55, 244-252. DOI: 10.1037/h0048480

Huguet, P., Latane, B., & Bourgeois, M. (1998). The emergence of a social representation of human rights via interpersonal communication: Empirical evidence for the convergence of two theories. *European Journal of Social Psychology*, 28, 831-846. DOI: 10.1002/(SICI)1099-0992

Jager, W., & Amblard, F. (2005). Uniformity, bipolarisation and pluriformity captured as generic stylized behaviour with an agent-based simulation model of attitude change. *Computational and Mathematical Organization Theory*, 10, 295-303. DOI: 10.1007/s10588-005-6282-2

Janis, I. L., Hovland, C. I., Field, P. B., Linton, H., Graham, E., Cohen, A. R., Rife, D., Abelson, R. P., Lesser, G. S., & King, B. T. (1959). *Personality and persuasibility.* New Haven: Yale University Press.

Kandel, D. B. (1978). Homophily, selection, and socialization in adolescent friendships. *American Journal of Sociology*, 84, 427-436. **DOI**: 10.1086/226792

Katz, E., & Lazarsfeld, P. F. (1955). *Personal influence*. Glencoe: Free Press.

Kitts, J. A. (2003). Egocentric bias or information management? Selective disclosure and the social roots of norm misperception. *Social Psychology Quarterly*, 66, 222-237. DOI: 10.2307/1519823

Kitts, J. (2006). Social influence and the emergence of norms amid ties of amity and enmity. *Simulation Modelling Practice and Theory*, 14, 407-422. DOI: 10.1016/j.simpat.2005.09.006

Krizan, Z., & Baron, R. S. (2007). Group polarization and choice-dilemmas: How important is self-categorization? *European Journal of Social Psychology*, 37, 191-201. DOI: 10.1002/ejsp.345

Lazarsfeld, P. F., & Merton, R. K. (1954). Friendship and social process: A substantive and methodological analysis. In M. Berger, T. Abel, & C. H. Page (Eds.), *Freedom and control in modern society* (pp. 18-66). New York, Toronto, London: Van Nostrand.

Lemaine, G. (1974). Social differentiation and social originality. *European Journal of Social Psychology*, 4, 17-52. DOI: 10.1002/ejsp.2420040103

Levendusky, M. S. (2009). The microfoundations of mass polarization. *Political Analysis*, 17, 162-176. DOI: 10.1093/pan/mpp003

Mackie, D. M. (1986). Social identification effects in group polarization. *Journal of Personality and Social Psychology*, 50, 720-728. DOI: 10.1037/0022-3514.50.4.720

Macy, M. W., Kitts, J., Flache, A., & Benard, S. (2003). Polarization in dynamic networks: A hopfield model of emergent structure. In R. Breiger, K. Carley, and P. Philippa (Eds.), *Dynamic social network modeling and analysis. Workshop summary and papers* (pp. 162-73). Washington D.C.: National Academy Press.

Mark, N. (1998). Beyond individual differences: Social differentiation from first principles. *American Sociological Review*, 63, 309-330. **DOI**: 10.2307/2657552

Mark, N. (2003). Culture and competition: Homophily and distancing explanations for cultural niches. *American Sociological Review*, 68, 319-345. DOI: 10.2307/1519727

Mäs, M., Flache, A., & Helbing, D. (2010). Individualization as driving force of clustering phenomena in humans. *PLoS Computational Biology*, 6. DOI: 10.1371/journal.pcbi.1000959.

Mason, W. A., Conrey, F. R., & Smith, E. R. (2007). Situating social influence processes: Dynamic, multidirectional flows of influence within social networks. *Personality and Social Psychology Review*, 11, 279-300. DOI: 10.1177/1088868307301032

Mavrodiev, P., Tessone C. J., & Schweitzer, F. (2013). Quantifying the effects of social influence. *Scientific Reports,* 3, 13-60. DOI: 10.1038/srep01360

Mazen, R., & Leventhal, H. (1972). The influence of communicator-recipient similarity upon the beliefs and behavior of pregnant women. *Journal of Experimental Social Psychology*, 8, 289-302. DOI: 10.1016/0022-1031(72)90019-4

McPherson, M., Smith-Lovin, L., & Cook, J. M. (2001). Birds of a feather: Homophily in social networks. *Annual Review of Sociology*, 27, 415-444. DOI: 10.1146/annurev.soc.27.1.415

Mucchi-Faina, A., & Cicoletti, G. (2006). Divergence vs. ambivalence: effects of personal relevance on minority influence. *European Journal of Social Psychology*, 36: 91–104. DOI: 10.1002/ejsp.278

Mucchi-Faina, A., & Pagliaro, S. (2008). Minority influence: The role of ambivalence toward the source. *European Journal of Social Psychology*, 38, 612-623. DOI: 10.1002/ejsp.486

Murray, D. R., & Schaller, M. (2012). Threat(s) and Conformity Deconstructed: Perceived Threat of Infectious Disease and Its Implications for Conformist Attitudes and Behavior. *European Journal of Social Psychology*, 42: 180-188.

Nowak, A.; Szamrej, J., & Latané, B. (1990). From Private Attitude to Public Opinion: A Dynamic Theory of Social Impact. *Psychological Review,* 97, 362-376.

Perloff, R. M. (1993). *The Dynamics of persuasion*. Hillsdale: Lawrence Erlbaum Associates.

Pilkington, N. W., & Lydon, J. E. (1997). The relative effect of attitude similarity and attitude dissimilarity on interpersonal attraction: Investigating the moderating roles of prejudice and group membership. *Personality and Social Psychology Bulletin*, 23, 107-116. DOI: 10.1177/0146167297232001

Platow, M. J., Mills, D., & Morrison, D. (2000). The effects of social context, source fairness, and perceived self-source similarity on social influence: A self-categorisation analysis. *European Journal of Social Psychology*, 30, 69-81. DOI: 10.1002/(SICI)1099-0992(200001/02)30:1<69::AID-EJSP980>3.0.CO;2-R

Raudenbush, S., Bryk, T., & Congdon, R. (2004). *HLM 6. Hierarchical linear and nonlinear modeling.* Skokie: Scientific Software International, Inc.

Rosenbaum, M. E. (1986). The repulsion hypothesis: On the non-development of relationships. *Journal of Personality and Social Psychology*, 51, 1156-1166. DOI: 10.1037/0022-3514.51.6.1156

Rosenberg, M. J. (1960). An analysis of affective-cognitive consistency. In M. J. Rosenberg, C. I. Hovland, W. J. McGuire, R. P. Abelson, & J. W. Brehm (Eds.), *Attitude organization and change*. New Haven: Yale University Press.

Rydgren, J. (2004). The logic of xenophobia. *Rationality and Society*, 16, 123-148. DOI: 10.1177/1043463104043712

Salzarulo, L. (2006). A continuous opinion dynamics model based on the principle of meta-contrast. *Journal of Artificial Societies and Social Simulation*, 9, http://jasss.soc.surrey.ac.uk/9/1/13.html

Sampson, E. E., & Insko, C. A. (1964). Cognitive consistency and performance in the autokinetic situation. *Journal of Abnormal and Social Psychology*, 68, 184-192. DOI: 10.1037/h0041242

Schachter, S. (1951). Deviation, rejection, and communication. *Journal of Abnormal and Social Psychology*, 46, 190–207. DOI: 10.1037/h0062326

Schwartz, S. H., & Ames, R. E. (1977). Positive and negative referent others as sources of influence: A case of helping. *Sociometry*, 40, 12-21. DOI: 10.2307/3033541

Sherif, M., & Hovland, C. I. (1961). *Social judgement: Assimilation and contrast effects in communication and attitude change.* New Haven: Yale University Press.

Skvoretz, J. (2013). Diversity, integration, and social ties: Attraction versus repulsion as drivers of intra- and intergroup relations. *American Journal of Sociology*, 119(2): 486-517. DOI: 10.1086/674050

Smeaton, G., Byrne, D., & Murnen, S. K. (1989). The repulsion hypothesis revisited: Similarity irrelevance or dissimilarity bias? *Journal of Personality and Social Psychology*, 56, 54-59. DOI: 10.1037/0022-3514.56.1.54

Smith, E. R., & Conrey, F. R. (2007). Agent-based modeling: A new approach for theory building in social psychology. *Personality and Social Psychology Review*, 11: 87-104. DOI: 10.1177/1088868306294789

Snijders, T. A. B., & Bosker, R. J. (1999). Multilevel Analysis. An Introduction to Basic and Advanced Multilevel Modeling. London, Sage.

Stotland, E., Zander, A., & Natsoulas, T. (1961). Generalization of interpersonal similarity. *Journal of Abnormal and Social Psychology*, 62, 250-256. DOI: 10.1037/h0041445

van Knippenberg, A., & Wilke, H. (1988). Social categorization and attitude-change. *European Journal of Social Psychology*, 18, 395-406. DOI: 10.1002/ejsp.2420180503

Wagner, C. G. (1978). Consensus through respect: a model of rational group decision-making. *Philosophical Studies*, 34, 335–349. DOI: 10.1007/BF00364701

Willer, R., Kuwabara, K., & Macy, M. W. (2009). The false enforcement of unpopular norms. *American Journal of Sociology*, 115, 451-490. DOI: 10.1086/599250

# APPENDIX

## Issues used in the experiments

**Table A1**. Means and standard deviations (in parentheses) of original opinions (O) and saliences (S) for issues used in the experiments

| Issues | Experiment 1 (N=89) | | Experiment 2 (N=110) | |
|---|---|---|---|---|
| | O | S | O | S |
| 1. The warning signs on cigarette boxes should cover 0…100 percent of the box total surface. | 43.4 (29.5) | 2.04 (.78) | 44.5 (29.5) | 2.20 (.88) |
| 2. Smoking should be allowed at 0…100 percent of tables in café's. | 32.8 (27.5) | 1.71 (.57) | 23.6 (27.1) | 1.75 (.75) |
| 3. The introduction of the euro brings advantages and disadvantages to us. 0…100 percent of all effects are advantages. | 51.4 (23.3) | 2.16 (.74) | 55.6 (24.9) | 2.30 (.76) |
| 4. The government should subsidize public transport in 0...100 percent. | 69.8 (22.5) | 1.56 (.58) | 65.7 (22.4) | 1.65 (.60) |
| 5. A demonstration needs police protection. Organizers should pay 0…100% of the costs of this. | 51.1 (33.1) | 2.21 (.67) | 44.0 (34.0) | 2.35 (.67) |
| 6. 0…100 percent of immigrants who come to the Netherlands for economic reason should receive a residence permit. | 34.8 (29.6) | 1.83 (.63) | 36.3 (29.6) | 1.82 (.68) |
| 7. Foreigners who want a residence permit for the Netherlands should pay 0…100 percent of their integration courses and tests. | 43.0 (33.5) | 1.92 (.79) | 42.2 (35.2) | 1.96 (.81) |
| 8. 0...100 percent of the streets in the centre of Groningen should have security camera surveillance. | 32.0 (27.1) | 2.15 (.70) | 33.7 (31.1) | 2.17 (.78) |
| 9. Sport activities of students should be financed by the university in 0…100% of total costs. | 49.9 (26.3) | 2.18 (.67) | 52.2 (27.8) | 2.20 (.81) |
| 10. Universities should be financed in 0…100 percent by tuition fees. | 40.0 (22.5) | 1.91 (.67) | 36.7 (22.0) | 2.03 (.64) |
| 11. The final grade of the overall study should be determined in 0…100 percent by the result of the Master's thesis. | 41.9 (21.9) | 2.13 (.73) | 32.6 (20.3) | 2.35 (.71) |
| 12. Somebody who is caught to draw graffiti should pay a fine of 0...100 euro. | 49.9 (28.6) | 2.47 (.66) | 52.0 (28.1) | 2.52 (.62) |
| 13. The fine for not cleaning up after your dog making dirt on the street should be 0…100 euro. | 43.3 (31.5) | 2.28 (.77) | 38.6 (25.8) | 2.25 (.64) |
| 14. The CEO of an industrial company has a limited budget for building a new plant. This budget has to be divided between employing additional employees | 48.4 (20.5) | 1.88 (.72) | 45.7 (20.7) | 1.83 (.59) |
| and investing in measures that protect the environment. 0…100 percent of the available budget should go to environment-protecting measures. | | | | |
| 15. The government has to divide an available budget between two options: building new highways or new high-speed railway tracks. 0…100 percent of these resources should be used to build new high-speed railway tracks. | 62.0 (20.4) | 1.88 (.64) | 65.8 (19.2) | 1.84 (.57) |
| 16. Schools in disadvantaged areas should receive 0…100 percent more financing than schools that are not in disadvantaged areas. | 33.0 (19.7) | 1.85 (.59) | 34.9 (21.2) | 1.80 (.54) |
| 17. The Dutch military has to divide 100 million euros between activities within the national borders, such as national defense and training, and missions outside the country. Foreign missions should receive 0…100 million euros. | 45.1 (20.2) | 2.22 (.65) | 45.4 (22.5) | 2.39 (.85) |
| 18. Students should pay 0...100 percent of the costs of language courses offered by the university. | 36.6 (25.0) | 2.29 (.79) | 33.7 (26.2) | 2.17 (.78) |
| 19. Foreign students should pay 0...100 percent of their Dutch language courses. | 42.8 (27.1) | 2.26 (.73) | 31.5 (28.3) | 2.17 (.73) |
| 20. Students should spend a maximum of 0…100 percent of the 40 hours weekly working time on paid work. | 41.2 (27.4) | 2.34 (.80) | 52.8 (32.4) | 2.54 (.81) |
| 21. In a family with two children, in which the husband works and the wife stays at home, the wife should take 0…100 percent of the household duties. | 71.9 (12.6) | 2.15 (.68) | | |
| 22. The local government faces two alternatives for extending the amount of housing: to build on non-residential areas or to rebuild, renovate or extend existing buildings. 0…100 percent of the resources should be devoted to build on non-residential areas. | 30.2 (21.0) | 2.03 (.73) | | |
| 23. The government should finance propagation of modern poetry in 0…100 percent of costs. | 20.4 (24.5) | 3.01 (.79) | | |
| 24. The government should pay 0...100 percent of the costs of child day-care. | 56.9 (25.0) | 1.73 (.56) | | |
| 25. A new shopping center is built. The costs of building a street to the shopping center has to be paid in 0…100 percent by the shopping center. | 49.6 (26.8) | 2.51 (.62) | | |
| 26. A maximum of 0…100 percent of the total EU budget should be spent on agriculture. | 23.7 (15.1) | 2.27 (.75) | | |
| 27. Israel should pay 0…100 percent of the costs of | 56.5 | 1.93 | | |

| | | |
|---|---|---|
| rebuilding in Lebanon. | (28.4) | (.70) |
| 28. For the problems around integration in the Netherlands the responsibility goes in 0…100 percent to Muslims (0 = the responsibility is in 0% of Muslims, and in 100% is of non-Muslims). | 38.9 (26.7) | 1.84 (.88) |
| 29. A professor at the university should spend 0…100 percent of his or her working time on teaching (thus not on research and not on administrative duties). | 45.0 (21.2) | 2.27 (.73) |
| 30. The final grade of a subject should be determined in 0…100 percent by a result of a written exam. | 71.3 (17.9) | 2.02 (.71) |
| 31. Students should spend 0…100 percent of the 40 hours weekly working time on their study. | 64.3 (19.9) | 2.06 (.82) |

*Notes:* All issues were measured on a 0…100 percentage scale. Salience was measured on an ordinal scale: "How important…" with answer categories "very important"=1, "important"=2, "unimportant"=3, "very unimportant"=4. Formulations are independent back-translations from Dutch.